\documentclass[a4paper,11pt]{article}
\pdfoutput=1
\usepackage{jheppub}
\usepackage{braket}
\usepackage{bm}
\usepackage{subcaption}
\usepackage{todonotes}
\usepackage{wrapfig}
\usepackage{float}  

\allowdisplaybreaks

\title{Electronic Wichmann-Kroll correction to the energy levels of muonic hydrogen-like atoms}

\author{Roman N. Lee}
\author{and Alexey V. Popov}

\affiliation{Budker Institute of Nuclear Physics, Novosibirsk 630090, Russia}

\emailAdd{r.n.lee@inp.nsk.su}
\emailAdd{al.v.popov@inp.nsk.su}

\abstract{Using modern multiloop calculation methods, we obtain correction to energy levels of muonic hydrogen-like atoms due to the leading-order ($\propto\alpha (Z\alpha)^3$) Wichmann-Kroll electronic potential. Our result is expressed in terms of generalized polylogarithms. As a byproduct, we also obtain the Wichmann-Kroll potential in momentum space in terms of the same functions. 
}

\begin{document}
	\maketitle
	\flushbottom
\section{Introduction}
	Muonic hydrogen-like atoms are bound systems of a muon and a nucleus. Since a muon is much heavier than an electron ($m_{\mu}/m_{e}\approx 207$), muonic atoms are much more sensitive to nuclear structure effects than ordinary electronic atoms. In particular, the Lamb shift $(2S - 2P)$ measurement in muonic hydrogen ($\mu^{-}p^{+}$) is currently the most accurate method for determining the proton charge radius. Experiments with this bound system provide accuracy an order of magnitude higher than those with ordinary hydrogen atom \cite{pohl2013muonic}.
	
	From theory side, muonic hydrogen-like atoms are traditional objects of study in quantum electrodynamics (QED). Both numerical and analytic computational techniques were used to obtain various corrections to the energy levels of these systems \cite{Borie1982,eides2007theory}. Modern multiloop methods, such as the IBP reduction and the differential equations method, extend the applicability of analytic approaches to a much wider class of problems. In particular, it concerns the problems related to the influence of light fermion (electron) vacuum polarization on the energy levels of muonic atoms.
	
\begin{wrapfigure}[10]{r}{0.35\textwidth}
	\centering
	\includegraphics[width=0.8\linewidth]{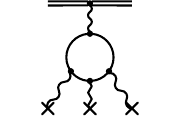}
	\caption{\small Feynman diagram for the Wichmann-Kroll correction.}
	\label{fig:WKdiag}
\end{wrapfigure}

	The important point in regard to the light particle vacuum polarization effect in 
	muonic atoms is that it gives contributions much larger than that of heavy particles \cite{pachucki1996theory}. This can be understood as follows. If the characteristic momentum transfer $q = m_\mu Z \alpha$ ($Z$ is nuclear charge) is much smaller than the mass $m$ of a particle in a loop $(q\ll m)$, the contribution is suppressed by the factor $\sim(q/m)^{2}= \left(m_\mu Z \alpha/m\right)^{2}$. In contrast, when the ratio $q/m= m_{\mu}Z\alpha/m_{e}$ is not small, which is the case for electron vacuum polarization in muonic atoms, the contribution is a nontrivial function of this ratio.	 

		In the present paper we derive an analytic expression for the electronic Wichmann-Kroll correction to the energy levels of muonic hydrogen-like atoms.\footnote{Previously this correction was calculated numerically in Refs. \cite{blomqvist1972vacuum,korzinin2018light}.} 	
		It corresponds to the diagram shown in Fig. \ref{fig:WKdiag}, where the upper double line corresponds to the bound muon, while the wavy lines with crosses denote the Coulomb field of the nucleus.
\section{Wichmann-Kroll correction to energy levels}
	We work in units $\hbar=c=1$ and set $m_{e}=1$ for simplicity. The Wichmann-Kroll (WK) potential $V_{WK}(r)$  \cite{wichmann1956vacuum} yields the following correction to the energy levels:
	\begin{equation}
		\delta E_{WK}=\int d^{3}\boldsymbol{r}|\psi_{nlm}(\boldsymbol{r})|^{2}V_{WK}(r),
		\label{eq:deltaEinit}
	\end{equation}
	where $\psi_{nlm}(\boldsymbol{r})$ is the Coulomb wave function. 
	\par Rewriting Eq. (\ref{eq:deltaEinit}) in momentum space, we obtain 
	\begin{equation}
		\delta E_{WK}=\int\frac{ d^{3}\boldsymbol{q}}{(2\pi)^{3}}\mathcal{F}_{nl} (|\boldsymbol{q}|)\widetilde{V}_{WK}(|\boldsymbol{q}|),
		\label{eq:deltaE}
	\end{equation}
	where $\mathcal{F}_{nl} (|\boldsymbol{q}|)= \int d^{3}\boldsymbol{r}\frac{\sin(|\boldsymbol{q}|r)}{|\boldsymbol{q}|r}|\psi_{nlm}(\boldsymbol{r})|^{2}$ is an angle-averaged Fourier transform of $|\psi_{nlm}(\boldsymbol{r})|^{2}$, 	
	and $\widetilde{V}_{WK}(|\boldsymbol{q}|)$ is the Fourier transform of the WK potential given by (see Fig. \ref{fig:WKdiag}):
	\begin{flalign}
		\widetilde{V}_{WK}(|\boldsymbol{q}|)&=\frac{iZ^{3}(4\pi\alpha)^{4}}{-q^{2}}\int\frac{d^{4}p}{(2\pi)^{4}}\frac{d^{3}\boldsymbol{q}_{1}}{(2\pi)^{3}}\frac{d^{3}\boldsymbol{q}_{2}}{(2\pi)^{3}}\notag\\
		&\times\frac{\operatorname{Tr}[\gamma_{0}(\widehat{p}_{1}+1)\gamma_{0}(\widehat{p}_{2}+1)\gamma_{0}(\widehat{p}_{3}+1)\gamma_{0}(\widehat{p}_{4}+1)]}{(1-p_{1}^{2})(1-p_{2}^{2})(1-p_{3}^{2})(1-p_{4}^{2})(-q_{1}^{2})(-q_{2}^{2})(-q_{3}^{2})}.\label{eq:WKpotentialasintegral}
	\end{flalign}
	Here $p_{1}=p+q_{1}$, $p_{2}=p$, $p_{3}=p-q_{2}$, $p_{4}=p-q_{2}-q_{3}$, $q=(0,\boldsymbol{q})$, $q_i=(0,\boldsymbol{q}_i)$,  $q_3=q-q_{1}-q_{2}$ and we imply the $-i0$ prescription for denominators.
	\par Substituting the non-relativistic expression for $\psi_{nlm}(\boldsymbol{r})$ from Ref. \cite{landau1958quantum}, we find
	\begin{multline}
		\mathcal{F}_{nl} (|\boldsymbol{q}|)=
		\frac{1}{2n \kappa}\frac{\binom{n+l}{2l+1}}{(2l+1)!}
		\int_{0}^{\infty}dx x^{2l+1}e^{-x}\sin\left(\kappa x\right) \big[{}_{1}F_{1}(-n+l+1,2l+2,x)\big]^{2}\\
		=\frac{\mathrm{Im}\left[(1+i\kappa)^{2n}\right]/\kappa  \, }{2n(1+\kappa^{2})^{2n}}\,{}_2F_1\left(-n-l,l+1-n;1;-\kappa^2\right).
	\end{multline}
	where $\kappa=\dfrac{|\boldsymbol{q}|n}{2m_r Z\alpha}$ with $m_r$ being the reduced mass of the muon-nucleus system.
	One can easily see that $\mathrm{Im}\left[(1+i\kappa)^{2n}\right]/\kappa$ is a polynomial in $\kappa^{2}$ of degree $n-1$ and ${}_{1}F_{1}$ is that of degree $n-l-1$. Thus, the function $\mathcal{F}_{nl} (|\boldsymbol{q}|)$ has the following form:
	\begin{equation}
		\mathcal{F}_{nl} (|\boldsymbol{q}|)=\sum_{k=l+1}^{2n-1}c_{k}\frac{1}{D^{k+1}}=\sum_{k=l+1}^{2n-1}\frac{c_{k}}{k!}\left(-\frac{\partial}{\partial\,(4\tau^{2})}\right)^{k}\frac{1}{D},\,\,\, D=4\tau^{2}+\boldsymbol{q}^{2},
		\label{eq:psi2ofD}
	\end{equation}
	where the coefficients $c_k$ do not depend on $\boldsymbol{q}^2$ and we have introduced
	\begin{equation}
		\tau=\frac{m_r Z\alpha}{m_e n}=\frac{m_r Z\alpha}{n}.
	\end{equation}
	
	Substituting (\ref{eq:psi2ofD}) into (\ref{eq:deltaE}), we obtain
	\begin{equation}
		\delta E_{WK}=\sum_{k=l+1}^{2n-1}\frac{c_{k}}{k!}\frac{d^{k}W(\xi)}{d\,\xi^{k}}\bigg|_{\xi=-4\tau^{2}},
		\label{eq:deltaEasfourloopint}
	\end{equation}
	where 	
	\begin{equation}
			W(\xi)= \int\frac{d^{3}\boldsymbol{q}}{(2\pi)^{3}}\,\frac{\widetilde{V}_{WK}(|\boldsymbol{q}|)}{-\xi+\boldsymbol{q}^{2}}.
			\label{eq:W}
	\end{equation}
\section{Calculation}
	Note that $W(-4\tau^2)$ in Eqs. \eqref{eq:deltaEasfourloopint} and \eqref{eq:W} (taking Eq. \eqref{eq:WKpotentialasintegral} into account) has the form of a four-loop Feynman integral. The denominator $4\tau^{2}-q^{2}$, which arises from $\mathcal{F}_{nl}$ (\ref{eq:psi2ofD}), is identical to that of the propagator for a particle of mass $m=2\tau$. Thus, we can compute the integral $W$ using modern multiloop calculation methods. 
	\par First, we replace
	\begin{equation}
		d^{3}\boldsymbol{q}_{1}\,d^{3}\boldsymbol{q}_{2}\,d^{3}\boldsymbol{q}\to d^{4}q_{1}\,d^{4}q_{2}\,d^{4}q_{3}\,\delta(q_{1}\cdot e_0)\delta(q_{2}\cdot e_0)\delta(q_{3}\cdot e_0),
	\end{equation}
	where $ e_0=(1,0,0,0)$.
	
	Then we introduce dimensional regularization in $d=4-2\epsilon$ dimensions. The denominators depending on $q$ (namely, $4\tau^2 - q^2$ and $q^2$) can be separated using the following relation:
	\begin{equation}
		\frac{1}{(4\tau^{2}-q^{2})q^{2}}=\frac{1}{4\tau^{2}(4\tau^{2}-q^{2})}+\frac{1}{4\tau^{2}q^{2}}
		\label{eq:tauparts}
	\end{equation}
	The first term in the right-hand side corresponds to the four-loop integral family
	\begin{equation}
		I(n_{1},\dots,n_{14})=\int d^{d}p\,d^{d}q_{1}\,d^{d}q_{2}\,d^{d}q_{3} \prod_{k=1}^{11} D_{k}^{-n_{k}}\,\prod_{l=12}^{14}\frac{\delta^{(n_{l}-1)}(-D_{l})}{(n_{l}-1)!},
		\label{eq:family}
	\end{equation}
	with
	\begin{flalign}
		D_{1-14}=&\big\{4\tau^{2}-(q_{1}+q_{2}+q_{3})^{2},\,\,\, 1-(p+q_{1})^{2},\,\,\,1-p^{2},\,\,\, 1-(p-q_{2})^{2},\,\,\,
		1-(p-q_{2}-q_{3})^{2},\notag\\
		&-q_{1}^{2},\,\,\, -q_{2}^{2},\,\,\, -q_{3}^{2},\,\,\, p\cdot e_0,\,\,\,p\cdot q_{3},\,\,\,q_{1}\cdot q_{2},\,\,\, q_{1}\cdot  e_0,\,\,\, q_{2}\cdot  e_0,\,\,\, q_{3}\cdot  e_0\big\}.\label{eq:denomsets}
	\end{flalign}
	Here $D_{9-11}$ are irreducible numerators and $D_{12-14}$ are cut denominators.
	
	The second term in Eq. (\ref{eq:tauparts}) corresponds to the naive limit $\tau\to0$ of the integrals (\ref{eq:family}).
	
\begin{figure}[h]
		\centering
		\includegraphics[width=0.9\linewidth]{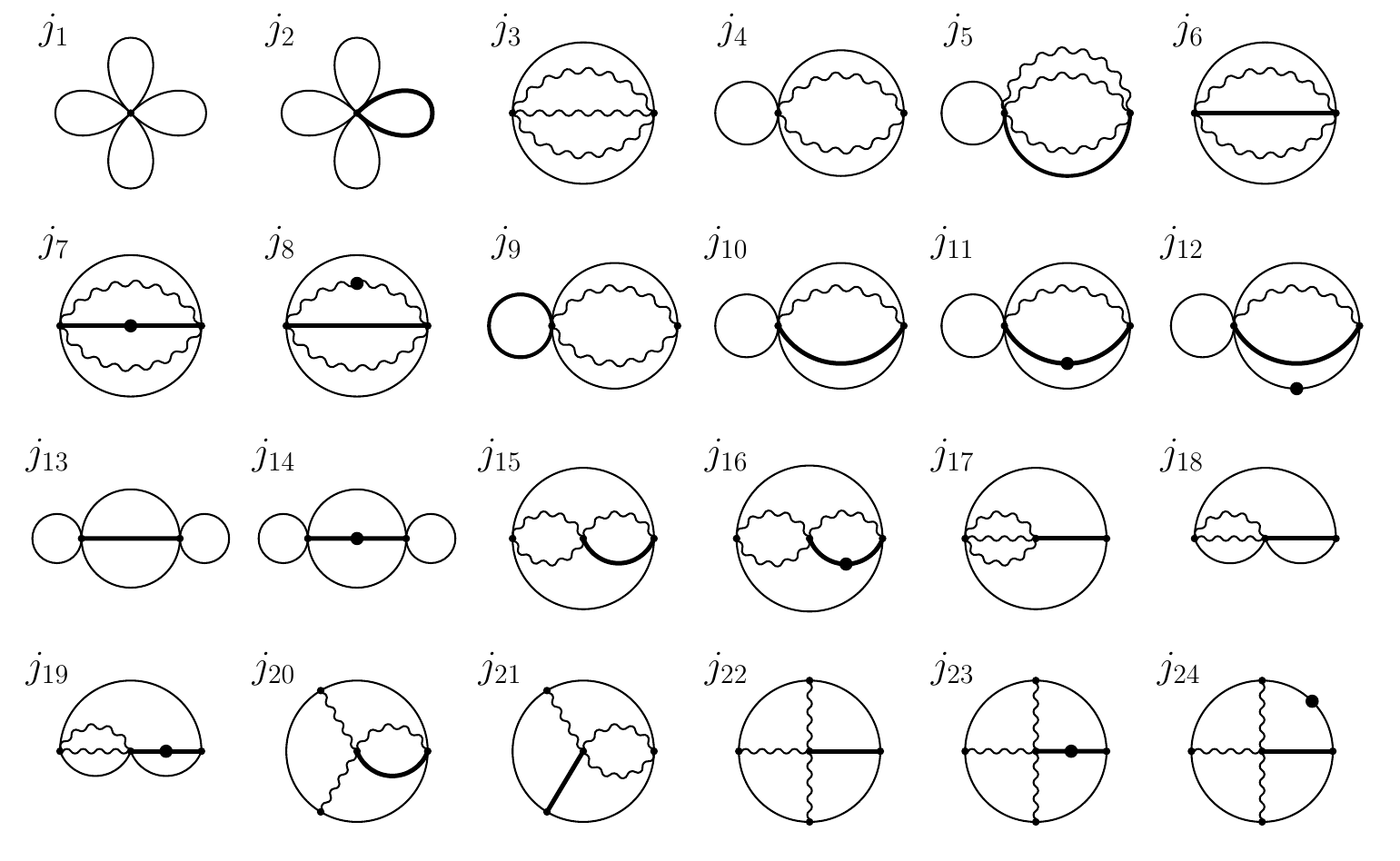}
		\caption{Master integrals; thin, thick and wavy lines denote denominators $1-p_{0}^{2}+\boldsymbol{l}^{2},\,\, 4\tau^{2}+\boldsymbol{l}^{2}$ and $\boldsymbol{l}^{2}$, respectively ($\boldsymbol{l}$ is a momentum corresponding to a line). Black dots denote squared denominator.}
		\label{fig:diagrams}
	\end{figure}

	Performing the IBP reduction with \texttt{LiteRed2} \cite{lee2013litered}, we reveal 24 master integrals (MIs). All MIs have indices $n_{9-11}=0$ and $n_{12-14}=1$. Therefore, they can be represented as
	\begin{equation}
		j_{k}=I(n_{1},\dots,n_{8})=\int_{-\infty}^{\infty}dp^{0}\bigg[\int d^{d-1}\boldsymbol{p}d^{d-1}\boldsymbol{q}_{1}d^{d-1}\boldsymbol{q}_{2}d^{d-1}\boldsymbol{q}_{3}\prod_{i=1}^{8}\widetilde{D}_{i}^{-n_{i}}\bigg],
		\label{eq:newmisform}
	\end{equation}
	where $\widetilde{D}_{i}=D_{i}|_{q_k^{0}=0}$. Then we can associate a diagram with the integral the square brackets of Eq. (\ref{eq:newmisform}). 
	The diagrams corresponding to MIs are shown in Fig. \ref{fig:diagrams}. 
	Using the IBP reduction, we obtain the differential system
	\begin{equation}
		\frac{d}{d\tau}\boldsymbol{j}=\mathcal{M}(\tau,\epsilon)\cdot\boldsymbol{j},
		\label{eq:dsgeneral}
	\end{equation}
	where $\boldsymbol{j}=(j_1,\ldots,j_{24})^\intercal$ is a column of master integrals.
	We solve this differential system by reducing it to the $\epsilon$-form using \texttt{Libra}, Refs. \cite{lee2015reducing,lee2021libra}. In order to find the $\epsilon$-form, we had to pass to the variable $x$, related to $\tau$ as follows:
	\begin{equation}
		\tau=\dfrac{2x}{1+x^{2}},\qquad x=\dfrac{1-\sqrt{1-\tau^{2}}}{\tau}.
	\end{equation}
	
	The resulting system has the following form
	\begin{equation}
		\partial_x \boldsymbol{J}=\epsilon \sum_{k=1}^{4} S_kw_k(x)\boldsymbol{J},
	\end{equation}
	where $S_k$ are constant matrices and 
	\begin{equation}
		w_{1}(x)=\frac{1}{x},\,\, w_{2}(x)=\frac{2x}{x^{2}+1},\,\, w_{3}(x)=\frac{2}{x^{2}+1},\,\, w_{4}(x)=\frac{2x}{x^{2}-1}.
		\label{eq:ws}
	\end{equation}
Consequently, the general solution is expressed in terms of the iterated integrals  
\begin{equation}
	\mathbb{I}(v_{1},v_{2},\dots,v_{n}|x)=\int_{0}^{x} dx_{1}\, v_{1}(x_{1}) \int_{0}^{x_{1}} dx_{2}\,v_{2}(x_{2})\,\,\dots\int_{0}^{x_{n-1}}dx_{n}\, v_{n}(x_{n}),
	\label{eq:II}
\end{equation}
	where $v_k\in\{w_1,w_2,w_3,w_4\}$.
	
	We fix the boundary conditions by calculating the following asymptotic coefficients:
	\begin{flalign}
	&[j_{1}]_{x^{0}},[j_{2}]_{x^{1-2\epsilon}},[j_{3}]_{x^{0}},[j_{4}]_{x^{0}},[j_{5}]_{x^{1-6\epsilon}},[j_{6}]_{x^{0}},[j_{6}]_{x^{2-6\epsilon}},[j_{6}]_{x^{1-2\epsilon}},[j_{9}]_{x^{1-2\epsilon}},\notag\\
	&[j_{10}]_{x^{0}},[j_{10}]_{x^{4-8\epsilon}},[j_{10}]_{x^{1-4\epsilon}},[j_{13}]_{x^{4-8\epsilon}},[j_{13}]_{x^{-2\epsilon}},[j_{15}]_{x^{0}},[j_{15}]_{x^{-6\epsilon}},[j_{17}]_{x^{-6\epsilon}},\notag\\
	&[j_{18}]_{x^{2-8\epsilon}},
	[j_{18}]_{x^{-2\epsilon}},[j_{20}]_{x^{-1-6\epsilon}},[j_{21}]_{x^{-1-6\epsilon}},[j_{22}]_{x^{0}},[j_{22}]_{x^{-2-8\epsilon}},[j_{22}]_{x^{-2-6\epsilon}},\label{eq:conditions}
	\end{flalign}
	where $[j_{k}]_{x^{\nu}}$ denotes the coefficient in front of $x^{\nu}$ in the asymptotic expansion of $j_{k}$ at $x\to 0$. In order to calculate these coefficients, 
	we use the Lee-Pomeransky representation \cite{lee2013critical} for the expression in square brackets in Eq. \eqref{eq:newmisform}. Then we obtain
	\begin{equation}
		j_{k}\propto\int_{-\infty}^{\infty}dp^{0}\intop_{\mathbb{R}_{+}^{M}}dy\, y^{n-1}G^{-(d-1)/2},
	\end{equation}
	where  $M$ is the number of denominators and $G=-(p^{0})^{2}A(y)+B(y)-i0$. Integrating over $p^{0}$, we obtain:
	\begin{equation}
		j_{k}\propto\intop_{\mathbb{R}_{+}^{M}}dy\, y^{n-1}\, A(y)^{-1/2}\, B(y)^{1 - d/2}.
	\end{equation} 
	\par Using the expansion-by-regions technique \cite{beneke1998asymptotic,pak2011geometric}, we determine the nonzero coefficients among (\ref{eq:conditions}) in terms of $\Gamma$-functions. In such an approach, however, it appears to be difficult to calculate the coefficient $[j_{22}]_{x^{0}}$. Therefore, we find this coefficient from the constraints at another point $x=1$. Specifically, the expansion-by-regions technique shows that the coefficient $[j_{22}]_{(1-x)^{1-2\epsilon}}$ vanishes. Using the $\epsilon$-expansion of the evolution operator, this coefficient can be expressed as a linear combination of $[j_{22}]_{x^{0}}$ and other coefficients listed in Eq. (\ref{eq:conditions}). Solving this constraint, we obtain $[j_{22}]_{x^{0}}$ as $\epsilon$-expansion up to a sufficiently high power.
	

\begin{wrapfigure}[9]{r}{0.3\linewidth}
	\centering\vspace{-4ex}
	
	\includegraphics[width=0.8\linewidth]{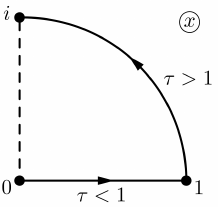}
	\caption{\small Path of $x$ when $\tau\in [0,\infty)$.}
	\label{fig:xpath}
\end{wrapfigure}
	Having the expression for $W$ in terms of the iterated integrals \eqref{eq:II}, we can obtain the expressions for the energy shifts from Eq. \eqref{eq:deltaEasfourloopint} using the differentiation formula
	\begin{equation}
		\frac{d}{dx}\mathbb{I}(v_{1},\dots,v_{n}|x)=v_{1}(x) \mathbb{I}(v_{2},\dots,v_{n}|x)\,.
	\end{equation}

	Note that the iterated integrals \eqref{eq:II} can be expressed in terms of Goncharov's polylogarithms \cite{goncharov1998multiple} with indices being fourth roots of unity or zero:
	\begin{equation}
		G(r_{1},\dots,r_{n}|x),\qquad r_{i}\in\{0,\pm i,\pm 1\}.
	\end{equation}
	When $x$ varies from $0$ to $1$, the original variable $\tau$ also varies from $0$ to $1$. For the region $\tau\in[1,\infty)$ we choose $x=e^{i\phi}$ with $\phi\in[0,\pi/2)$, see Fig. \ref{fig:xpath}. For this region it is convenient to express $\mathbb{I}$ via Goncharov's polylogs of the variable
	\begin{equation}
		z=i\frac{1-x}{1+x}=\sqrt{\frac{\tau-1}{\tau+1}}.
	\end{equation}
%
%

\section{Results and conclusion}
According to the results of the previous section, we have obtained the expression for $\delta E_{WK}^{nl}$ in terms of the derivatives of the function $W$ up to $(2n-1)$-th order. The function $W$ has been expressed via iterated integrals \eqref{eq:II}, see Appendix \ref{eq:WviaII}, or fourth root of unity Goncharov's polylogarithms.
We provide the main results of the present paper in the following supplementary  files:
\begin{itemize}
	\item \texttt{W.m} --- an expression for the integral $W$ in terms of iterated integrals.
	\item \texttt{IItoGx} --- substitution rules for $\mathbb{I}$ via $G$ of $x$.
	\item \texttt{IItoGz} --- substitution rules for $\mathbb{I}$ via $G$ of $z$.
	\item \texttt{main.nb} --- a program for obtaining $\delta E_{WK}^{nl}$ with arbitrary $n$ and $l$ and $\widetilde{V}_{WK}(Q)$.	
\end{itemize}
 In particular, the notebook \texttt{main.nb} contains a code which analytically calculates the energy shift $\delta E_{WK}^{nl}$ for any $(n,l)$ state.\footnote{Note that even for the ground state the expression for the energy shift is rather cumbersome and we do not present it in the text to save space.} 

\begin{figure}[H]
		\centering
		\includegraphics[width=1\linewidth]{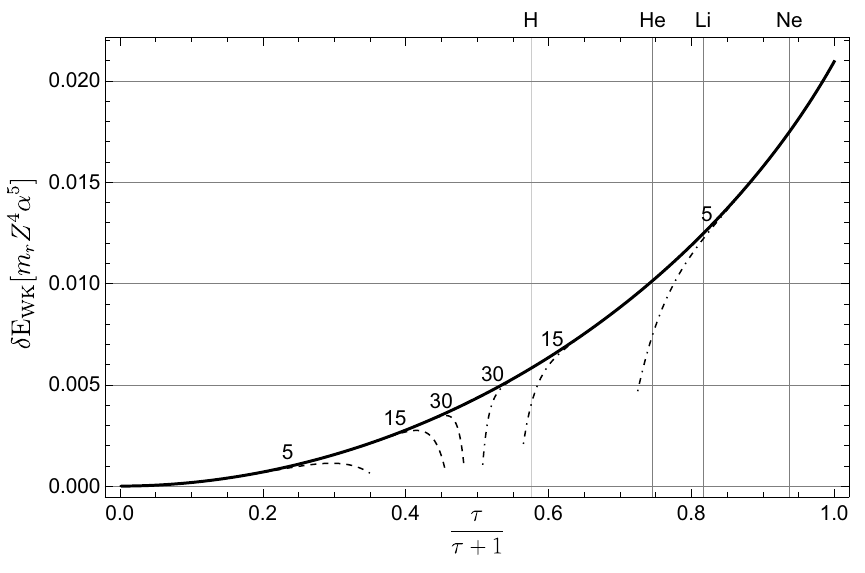}
		\caption{$\delta E_{WK}^{1s}$ in units of $m_{r}Z^{4}\alpha^{5}$ (solid line), asymptotic approximations at $\tau\to0$ (dashed lines) and at $\tau\to\infty$ (dash-dotted lines); numbers are the approximation orders. Labels $X$ denote the bound systems $\mu X$.}
		\label{fig:fn1}
	\end{figure}

\begin{figure}[H]
		\centering
		\includegraphics[width=1\linewidth]{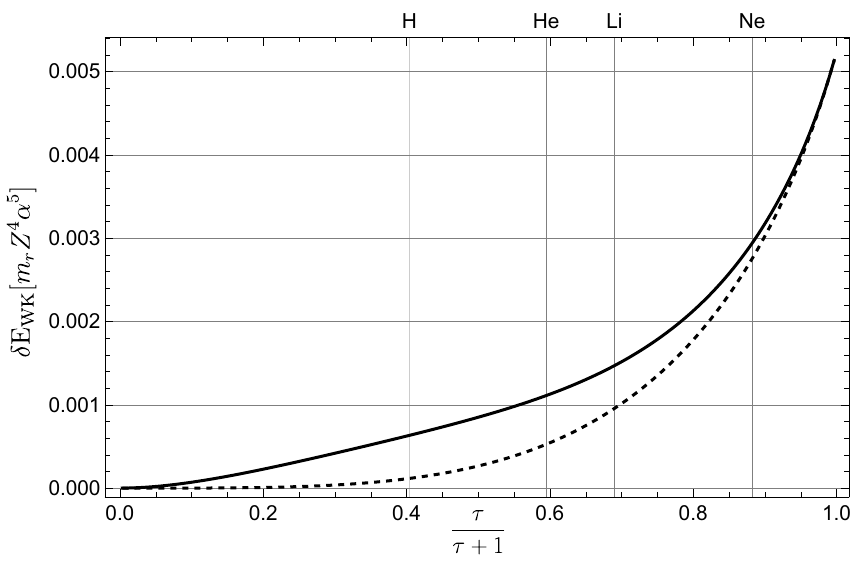}
		\caption{$\delta E_{WK}^{2s}$(solid line) and $\delta E_{WK}^{2p}$ (dashed line) in units of $m_{r}Z^{4}\alpha^{5}$.}
		\label{fig:fn2}
	\end{figure}

As an example of numerical results, in Figs. \ref{fig:fn1} and \ref{fig:fn2} we show $\delta E_{WK}^{nl}$ as function of $\tau=\dfrac{m_r Z\alpha}{m_{e}n}$ for $n=1$ and $2$, respectively. Our results for the $1s$, $2s$, and $2p$ states are consistent with the numerical results from Ref. \cite{korzinin2018light}.

Given our analytical result, it is easy to calculate the deep asymptotic expansions of the found energy shifts at small and large $\tau$. The leading terms of these expansions, 
	\begin{align}
		\delta E_{WK}^{nl}&\stackrel{\tau\to 0}{\simeq}m_{r}Z^{4}\alpha^{5}\left[\left(\frac{19}{45\pi}-\frac{\pi}{27}\right)\frac{\delta_{l,0}}{n}\tau^{2}+\left(\frac{1}{16}-\frac{31\pi^{2}}{2880}\right)\delta_{l,0}\tau^{3}+\mathcal{O}(\tau^{4})\right],\label{eq:tauzero}\\
		\delta E_{WK}^{nl}&\stackrel{\tau\to \infty}{\simeq}m_{r}Z^{4}\alpha^{5}\bigg[\left(\frac{\pi}{6}-\frac{7}{9\pi}-\frac{2\zeta(3)}{3\pi}\right)\frac{1}{n^{2}}+\left(2\zeta(3)-\frac{\pi^{2}}{4}\right)\frac{1}{n\tau}+\mathcal{O}(\tau^{-2})
		\bigg],\label{eq:tauinf}
	\end{align}
	agree with known results. In particular, the small-$\tau$ expansion agrees with Ref. \cite{mohr1983energy}. The large-$\tau$ expansion can be easily reproduced from the first two terms of the Wichmann-Kroll potential expansion, presented in Ref. \cite{blomqvist1972vacuum}.
	Note that the point $\tau=1$ is a singular point of the differential system for the master integrals. Therefore, although the function $W$, and, so, the energy shifts, are analytic at this point, the convergence radii of the expansions with respect to $\tau$ and $1/\tau$ are equal to $1$. To demonstrate this, we show in Fig. \ref{fig:fn1} the asymptotic expansions up to order 5, 15, and 30 with respect to $\tau$. One can clearly see that even deep expansions significantly depart from the exact curve in the vicinity of the point $\tau=1$ as expected.
	
	We emphasize that we have carried out our calculations with non-relativistic Coulomb wave functions. Therefore our results do not include relativistic corrections. It is especially important  when $m_r\sim m_{e}$, $\tau\sim Z\alpha\ll 1$ and $l\neq0$. In this case the contribution of relativistic corrections in $E_{WK}^{nl}$ is of the same order as the leading term $\propto\tau^4$ of the asymptotic expansion at $\tau\to 0$.
	 
	It is remarkable that the Fourier transform of the WK potential ($\widetilde{V}_{WK}(Q))$ can also be obtained from the expression for $W$. Indeed, the equation (\ref{eq:W}) can be represented as a dispersion relation for the function $W$:
	\begin{equation}
		W(\xi)=\frac{1}{\pi}\intop_0^{\infty}dQ^2 \frac{\tfrac{Q}{4\pi}\widetilde{V}_{WK}(Q)}{Q^2-\xi}.
	\end{equation} 
	Thus, $\widetilde{V}_{WK}(Q)$ is related to the imaginary part of $W$, evaluated at $\xi=Q^{2}$, as follows:
	\begin{equation}
		\widetilde{V}_{WK}(Q)=\dfrac{4\pi}{Q}\, \mathrm{Im}\left[W(Q^{2}+i0)\right]=\dfrac{2\pi}{iQ}\left[W(Q^{2}+i0)-W(Q^{2}-i0)\right].
	\end{equation}
	Given the expression for $W(-4\tau^{2})$ via the variable $x$ we can find $W(Q^2+i0)$ by substitution $x\to -i t$, where $t=\dfrac{\sqrt{4+Q^2}-2}{Q}$. We provide the implementation of this method for obtaining $\widetilde{V}_{WK}(Q)$ in the notebook \texttt{main.nb}.
	
	To summarize, we have obtained corrections to the energy levels of muonic atoms due to the electronic Wichmann-Kroll potential, as well as the Fourier transform of this potential expressed in terms of Goncharov's polylogarithms.

	\appendix
	\section{Expression for W}
	Quantity W is expressed in terms of iterated integrals defined in Eq. (\ref{eq:II}), as follows
	\begin{multline}
		W=
		\frac{(1+x^2) Z^3 \alpha ^4}{\pi} \bigg[\frac{\pi }{8}-\frac{\pi ^3 (9+21 x^2+8 x^4)}{576 x^2}+\frac{(1-x^2) (\mathbb{I}_{3}+2 (\pi \mathbb{I}_{4}+\mathbb{I}_{23}-2 \mathbb{I}_{43}))}{8 x^2}\\
		+\frac{(3 x^2-1) (\pi  \mathbb{I}_{2}-2 \mathbb{I}_{13})}{16 x^2}
		+\frac{18(\mathbb{I}_{2}+\pi  \mathbb{I}_{3}-2 \mathbb{I}_{33}-L)-14-3 \pi ^2}{72 x}\\
		+\frac{(15 x^2+8 x^4-3) (\pi  \mathbb{I}_{12}-2 \mathbb{I}_{113})}{48 x^2}
		+\frac{3+2 x^2}{6}  (2 \mathbb{I}_{113}+\mathbb{I}_{132}-L \mathbb{I}_{13})\\
		+\frac{(3-9 x^2-4 x^4) (\pi 
		\mathbb{I}_{14}+\mathbb{I}_{123}-2 \mathbb{I}_{143})}{12 x^2}+\frac{(4+9 x^2-9 x^4-4 x^6)
		(\mathbb{I}_{143}+\mathbb{I}_{413}+\mathbb{I}_{432}-L \mathbb{I}_{43})}{12 x^4}\\
		+\frac{(3+10 x^2+3 x^4) (\pi  (\mathbb{I}_{32}-2 \mathbb{I}_{34})+2 (\mathbb{I}_{133}-\mathbb{I}_{323}+\mathbb{I}_{332}+2 \mathbb{I}_{343}-L \mathbb{I}_{33}))}{12 x^3}\\
		-\frac{1+3 x^2-3 x^4-x^6}{12 x^4} (\pi  (\mathbb{I}_{22}-2 (\mathbb{I}_{24}+\mathbb{I}_{42}-2 \mathbb{I}_{44}))+2 (\mathbb{I}_{123}-\mathbb{I}_{223}+\mathbb{I}_{232}-L
		\mathbb{I}_{23}\\
		+2 (\mathbb{I}_{243}+\mathbb{I}_{413}+\mathbb{I}_{423}-2 \mathbb{I}_{443})))
		+\frac{(1+x^2) (\pi  (\pi  \mathbb{I}_{13}+12 \mathbb{I}_{114}-6 \mathbb{I}_{133})+12 (\mathbb{I}_{1123}-2
		\mathbb{I}_{1143}+\mathbb{I}_{1333}))}{48 x^2}\\
		+\frac{1+7 x^2+7 x^4+x^6}{16 x^4} (-\pi  \mathbb{I}_{122}+2 (\pi  (\mathbb{I}_{124}+\mathbb{I}_{142}-2 \mathbb{I}_{144})+\mathbb{I}_{1213}+\mathbb{I}_{1223}\\
		-2 (\mathbb{I}_{1243}+\mathbb{I}_{1413}+\mathbb{I}_{1423}-2 \mathbb{I}_{1443})))
		+\frac{1+5 x^2+5 x^4+x^6}{192 x^4} (\pi ^3 \mathbb{I}_{2}-12 (\pi  \mathbb{I}_{212}-2 L\, (\mathbb{I}_{123}-2 \mathbb{I}_{143}+\mathbb{I}_{213})\\
		+2 (-\pi  \mathbb{I}_{214}+2 \mathbb{I}_{1123}-4 \mathbb{I}_{1143}+2 \mathbb{I}_{1213}+\mathbb{I}_{1232}-2 \mathbb{I}_{1413}-2
		\mathbb{I}_{1432}+\mathbb{I}_{2113}-\mathbb{I}_{2123}+\mathbb{I}_{2132}+2 \mathbb{I}_{2143})))\\
		+\frac{1+2 x^2+x^4}{192 x^3} (\pi ^3 \mathbb{I}_{3}-6 \pi  \mathbb{I}_{312}+12 (2 L\, (\mathbb{I}_{313}-\mathbb{I}_{343})-2 \mathbb{I}_{1313}+2 \mathbb{I}_{1343}-3
		\mathbb{I}_{3113}\\
		+2(-\mathbb{I}_{3132}+\mathbb{I}_{3143}+\mathbb{I}_{3413}+\mathbb{I}_{3432})))\bigg]
		,
		\label{eq:WviaII}
	\end{multline}
	where $\mathbb{I}_{i_{1}\dots i_{k}}$ is the notation for $\mathbb{I}(w_{i_{1}},\dots,w_{i_{k}}|x)$ and $L=\log{4x}=\mathbb{I}_{1}+2\log{2}$.
\bibliographystyle{JHEP}
\bibliography{WK}
\end{document}